\documentclass[epsfig]{article}
\usepackage{amssymb}
\usepackage{graphicx}
\usepackage{amsmath}

\input epsf.sty
\input{tcilatex}
\begin{document}

\title{{\LARGE On} {\LARGE Braneworld Inflation Models in Light of WMAP7 data%
}}
\author{R.Zarrouki$^{1}$ Z.Sakhi$^{2}$ and M.Bennai$^{1,2}$\thanks{%
Corresponding authors: mdbennai@yahoo.fr,m.bennai@univh2m.ac.ma} \\
$^{1}${\small Laboratoire de Physique de la Mati\`{e}re Condens\'{e}e
(URAC10), }\\
\ {\small Facult\'{e} des Sciences Ben M'sik, B.P. 7955, Universit\'{e}
Hassan II-Mohammedia, \ \ }\\
\ {\small Casablanca, Maroc.}\\
$^{2}${\small Groupement National de Physique des Hautes Energies, Focal
point,}\\
\ \ {\small \ LabUFR-PHE, Rabat, Morocco.}}
\maketitle

\begin{abstract}
We are interested on studing various inflationary spectrum perturbation
parameters in the context of the \emph{Randall-Sandrum} type $2$ Braneworld
model. We consider in particular three types of potentials. We apply the
slow-roll approximation in the high energy limit to constraint the parameter
potentials by confronting our results to recent WMAP7 observations. We show
that, for some values of the e-folding number $N,$ the monomial potential
provides the best fit results to observations data.


\end{abstract}

\date{}
\tableofcontents

\newpage

\section{Introduction}

In the last few years, braneworld model\cite{Brax} has become a fundamental
paradigm of modern cosmology. One of the most used scenario is the \emph{%
Randall-Syndrum }type-2 model\cite{Randall}. In this theory,\emph{\ }our
four-dimensional Universe is considered as a $3$-brane embedded in
five-dimensional anti-de Sitter space-time ($AdS5$), while gravity can be
propagated in the bulk. The main characteristic feature of this model is
that the \emph{Friedmann} equation is modified by an additional term
proportional to quadratic energy density\cite{Binetruy}. Braneworld
inflation, was mainly shown to play a fondamental\ role in divers
cosmological investigations of early universe and has been proposed to solve
some problems of actual cosmological observations like as dark energy\cite%
{darkenergy}, tachyonic inflation\cite{taychons} or black holes systems\cite%
{BHRS}. Recently, some generalized potentials and supersymmetic versions
were studied in the framework of \emph{Randall-Syndrum }model and a best fit
to recent observations was given\cite{ZSB, BZCB}.

In the present work, we consider various types of potential models, in
particular the exponential, monomial and inverse power law. We have analyzed
in some details divers perturbation parameter spectrum and show that the
monomial potential provides the best fit results to more recent WMAP7 data.

In Sec.2, we begin by recalling the foundation of the braneworld inflation.
We have adopted here the slow-roll approximation and have considered the
high energy limit to derive various inflation observables. In Sec.3, we
present our work concerning a different models of potential in the framework
of the \emph{Randall-Syndrum} type 2 model and give a comparative study of a
exponential, monomial and inverse power law potentials. In the last section,
a conclusion and perspectives are given.

\section{Slow-roll approximation in braneworld inflation}

In this section, we start by recalling briefly some fundamentals of \emph{%
Randall-Sundrum} type-$2$ model. One of the most relevant consequences of
this model is the modification of the \emph{Friedmann} equation for energy
density of the order of the brane tension, and also the appearance of an
additional term, usually considered as dark radiation term. In the case
where the dark radiation term is neglected, the gravitational \emph{Einstein}
equations, leads to the modified \emph{Friedmann} equation on the brane as%
\cite{Binetruy}%
\begin{equation}
H^{2}={\frac{8\pi }{3m_{pl}^{2}}}\rho \left[ 1+{\frac{\rho }{2\lambda }}%
\right] ,
\end{equation}%
where $H$ is the \emph{Hubble} parameter, $\rho $ is the energy density, $%
\lambda $ is the brane tension and $m_{pl}$ is the \emph{Planck} mass. Note
that the crucial correction to standard inflation is given by the density
quadratic term $\rho ^{2}$.

Note also that in the limit $\lambda \rightarrow \infty ,$ we recover
standard four-dimensional general relativistic results. Moreover, in the
high energy limit i.e. $\rho \gg \lambda ,$ the dynamic of the universe in $%
5 $ dimensions will be governed by the simplest equation given by%
\begin{equation}
H^{2}=\frac{4\pi }{3m_{pl}^{2}}\frac{\rho ^{2}}{\lambda }.
\end{equation}%
On the other hand, the matter in $3$-brane is dominated by a scalar field
with energy density of the form $\rho =\frac{\dot{\phi}^{2}}{2}+V\left( \phi
\right) ,$ where $V(\phi )$ is the scalar field potential responsible of
inflation.

Along with these equations, one also has a second inflation \emph{%
Klein--Gordon} equation governing the dynamic of the scalar field as%
\begin{equation}
\ddot{\phi}+3H\dot{\phi}+V^{\prime }=0,
\end{equation}%
where $\dot{\phi}=\frac{\partial \phi }{\partial t}$, $\ddot{\phi}=\frac{%
\partial ^{2}\phi }{\partial t^{2}}$ and $V^{\prime }=\frac{dV\left( \phi
\right) }{d\phi }.$

This is a second-order evolution equation which follows from conservation
condition of energy--momentum tensor $T_{\mu \nu }$. To calculate some
physical quantities as scale factor or perturbation spectrum, one has to
solve Eqs.(1,3) for some specific potentials $V(\phi )$. To do so, the
slow-roll approximation was introduced and applied by many authors to derive
perturbation spectrum of inflation\cite{Liddle}. \bigskip In this work, we
apply slow-roll approximation ( $\dot{\phi}^{2}\ll $ $V\left( \phi \right) $
and $\ddot{\phi}\ll $ $H\dot{\phi}$ ) and we use the well known slow-roll
parameters\cite{Maartens}, to calculate perturbation spectrum. 
\begin{equation}
\epsilon =\frac{m_{pl}^{2}}{4\pi }\frac{\lambda V^{\prime 2}}{V^{3}},\text{
\ \ \ \ \ }\eta =\frac{m_{pl}^{2}}{4\pi }\frac{\lambda V^{\prime \prime }}{%
V^{2}},
\end{equation}%
where $V^{\prime \prime }=\frac{d^{2}V}{d\phi ^{2}}.$ We signal that, during
inflation we have the following conditions%
\begin{equation}
\epsilon \ll 1,\text{ \ \ \ \ }\mid \eta \mid \ll 1.
\end{equation}%
The small quantum fluctuations in the scalar field lead to fluctuations in
the energy density which was studied in a perturbative theory\cite{Lyth}. As
discussed in\cite{Kazuya} quantum fluctuations effect of the inflaton are
generally negligibles, since the coupling of the scalar field to bulk
gravitational fluctuations only modifies the usual 4D predictions at the
next order in the slow-roll expansion. So, one can define the power spectrum
of the curvature perturbations as%
\begin{equation}
P_{R}\left( k\right) =\left( \frac{H^{2}}{2\pi \overset{\cdot }{\phi }}%
\right) ^{2},
\end{equation}%
where $k$ is the wave number.

In relation to $P_{R}\left( k\right) $, the scalar spectral index is defined
as\cite{Lyth} 
\begin{eqnarray}
n_{s}-1 &=&\frac{d\ln P_{R}\left( k\right) }{d\ln k}, \\
&=&-6\epsilon +2\eta .
\end{eqnarray}%
On the other hand, the quantum fluctuations in the scalar field lead also to
fluctuations in the metric. In this way, one can define the amplitude of
tensor perturbations as\cite{Langlois}%
\begin{equation}
P_{g}\left( k\right) =\frac{64\pi }{m_{pl}^{2}}\left( \frac{H}{2\pi }\right)
^{2}F^{2}\left( x\right) ,
\end{equation}%
where $x=Hm_{pl}\sqrt{\frac{3}{4\pi \lambda }}$ and $F^{2}\left( x\right)
=\left( \sqrt{1+x^{2}}-x^{2}\sinh ^{-1}\left( \frac{1}{x}\right) \right)
^{-1}$. Note that in the high-energy limit ($V\gg \lambda $), $F^{2}\left(
x\right) \approx \frac{3}{2}x=\frac{3}{2}\frac{V}{\lambda }.$

These results lead to the ratio of tensor to scalar perturbations $r$ 
\begin{equation}
r=\frac{P_{g}\left( k\right) }{P_{R}\left( k\right) }.
\end{equation}%
As function of $\epsilon $, the inflation parameter $r$ become%
\begin{equation}
r=24\epsilon .
\end{equation}%
Other perturbation quantity is the running of the scalar index $\frac{dn_{s}%
}{d\ln k},$ which is given in terms of $V\left( \phi \right) $ as%
\begin{eqnarray}
\frac{dn_{s}}{d\ln k} &=&\frac{m_{pl}^{2}}{2\pi }\frac{V^{\prime }\lambda }{%
V^{2}}(3\frac{\partial \varepsilon }{\partial \phi }-\frac{\partial \eta }{%
\partial \phi }),  \notag \\
&=&-\frac{m_{pl}^{4}\lambda ^{2}}{8\pi ^{2}}(9\frac{V^{\prime 4}}{V^{6}}-8%
\frac{V^{\prime \prime }V^{\prime 2}}{V^{5}}+\frac{V^{\prime \prime \prime
}V^{\prime }}{V^{4}}),
\end{eqnarray}%
where $V^{\prime \prime \prime }=\frac{d^{3}V}{d\phi ^{3}}.$

Another important caracteristic inflationary parameter is the number of
e-folding $N$ defined by\cite{Maartens}%
\begin{equation}
N=-\frac{4\pi }{\lambda m_{pl}^{2}}\int_{\phi _{\ast }}^{\phi _{end}}\frac{%
V^{2}}{V^{\prime }}d\phi ,
\end{equation}%
where $\phi _{\ast }$ and $\phi _{end}$ are the values of the scalar field
at the epoch when the cosmological scales exit the horizon and at the end of
inflation, respectively.

In the next section, we give our results concerning a three potential models
and present the evolution of various inflationary perturbation spectrum
according to different potential parameters.

\section{Pertubation spectrum for various potentials}

\subsection{Exponential model}

The exponential potentiel was studied in various occasions, for example the
authors in ref.\cite{Copeland} have shown that inflation becomes possible in
Braneworld model for a class of potentials ordinarily too steep to sustain
accelerated expansion. They have also shown that this potentiel allows a
particularly natural implementation of reheating via gravitational particle
production.

Here we consider an exponential potential of \ following type%
\begin{equation}
V=V_{0}\exp \left( -\frac{\alpha }{m_{p}}\phi \right) ,
\end{equation}%
where $V_{0}$ and $\alpha $ are constants.

For this type of potential, the scalar spectral index $n_{s}$ and the ratio
of tensor to scalar perturbations $r$ take respectively the following
expressions%
\begin{eqnarray}
n_{s}-1 &=&-\frac{\alpha ^{2}\lambda }{\pi }\frac{1}{V}, \\
r &=&\frac{6\alpha ^{2}\lambda }{\pi }\frac{1}{V}.
\end{eqnarray}%
The running of the scalar index is presented by%
\begin{equation}
\frac{dn_{s}}{d\ln k}=-\frac{\alpha ^{4}\lambda ^{2}}{4\pi ^{2}V^{2}}.
\end{equation}%
Inflation ends when $\epsilon =1$, the equation $\epsilon =\frac{\alpha
^{2}\lambda }{4\pi V}$ leads to%
\begin{equation}
V_{end}=\frac{\alpha ^{2}\lambda }{4\pi }.
\end{equation}%
Taking into account the equation$\ $(13), the value of the potential when
the cosmological scales exit the horizon is 
\begin{equation}
V_{\ast }=\frac{\alpha ^{2}\lambda }{4\pi }\left( N+1\right) .
\end{equation}%
In terms of $N$ the inflation parameters become%
\begin{eqnarray}
n_{s}-1 &=&-\frac{4}{N+1}, \\
r &=&\frac{24}{N+1}, \\
\frac{dn_{s}}{d\ln k} &=&-\frac{4}{\left( N+1\right) ^{2}}.
\end{eqnarray}%
By using the equations (20) and (21) and WMAP7 observations data\cite{WMAP7},%
\begin{eqnarray}
0.963 &\leqslant &n_{s}\leqslant 1.002\text{\ \ }(95\%CL), \\
r &<&0.36\text{ \ \ }(95\%CL),
\end{eqnarray}%
we can show that $N>107.$ Thus, we can deduce an upper limit for the
e-folding number without running.

The power spectrum of the curvature perturbations is given by%
\begin{equation}
P_{R}=\frac{\lambda \alpha ^{6}\left( N+1\right) ^{4}}{48\pi ^{3}m_{pl}^{4}}.
\end{equation}%
The observed value for $P_{R}$ from WMAP7 is 
\begin{equation}
P_{R}=\left( 2.28\pm 0.15\right) \times 10^{-9}\text{ }(95\%CL).
\end{equation}%
According to Eq.(26), the condition $N>107$ implies that%
\begin{equation}
\lambda \lesssim \frac{2.65\times 10^{-14}}{\alpha ^{6}}m_{pl}^{4},
\end{equation}%
which constitutes the necessary and sufficient condition on the brane
tension with respect to parameter $\alpha $ so that the exponential
potential can describe the early inflation of the universe.

\subsection{Monomial model}

The exponential potential was recently introduced to describe tachyonic
inflation, but in standard inflation this potential cannot reproduce many
phenomena such dissipation. The chaotic potential and it's generalisation%
\cite{Sanchez}, can occur for field values, when the cosmological scales
exit the horizon $\phi _{\ast }$, below the four-dimensional $Planck$ scale
i.e. $\phi _{\ast }<m_{pl}$. In ref.\cite{ZSB}, we have studied a more
genaralized version of this potential in Branewold model, and shown that the
observation bound are satisfied. On the other hand, liddle et al.\cite%
{Liddle} have shown that observational constraints can be respected for an
monomial potential, in particular for $n=2$. Here, we are interested on
studing the variation of various inflationary parameters, in Braneworld
scenario, as function of $n$ for divers e-folding number values.

In this work we consider a potential of the form%
\begin{equation}
V=M\phi ^{n},
\end{equation}%
where $n$ is constant and $M$ is a parameter of dimension $\left[ E\right]
^{4-n}.$

For this type of potential, the scalar spectral index $n_{s}$ and the ratio $%
r$ are respectively given by%
\begin{eqnarray}
n_{s}-1 &=&-\frac{m_{pl}^{2}\lambda }{2\pi M}\frac{n}{\phi ^{n+2}}\left(
2n+1\right) , \\
r &=&\frac{6m_{pl}^{2}\lambda }{\pi M}\frac{n^{2}}{\phi ^{n+2}}.
\end{eqnarray}%
The running of the scalar index is presented by%
\begin{equation}
\frac{dn_{s}}{d\ln k}=-\frac{m_{pl}^{4}\lambda ^{2}n^{2}\left( n+2\right)
\left( 2n+1\right) }{8\pi ^{2}M^{2}\phi ^{2n+4}}.
\end{equation}%
To express the previous parameters of inflation only in terms of\ the
e-folding number $N$ and the exponent $n$, it is convenient to calculate the
values of $\phi _{end}$ and $\phi _{\ast }$ using Eq.(13)%
\begin{eqnarray}
\phi _{end}^{n+2} &=&\frac{m_{pl}^{2}\lambda }{4\pi M}n^{2}, \\
\phi _{\ast }^{n+2} &=&\frac{m_{pl}^{2}\lambda }{4\pi M}n\left( n\left(
N+1\right) +2N\right) .
\end{eqnarray}%
The power spectrum of the curvature perturbations is given by%
\begin{equation}
P_{R}=\frac{16\pi \lambda ^{\frac{n-4}{n+2}}M^{\frac{6}{n+2}}m_{pl}^{\frac{%
2n-8}{n+2}}}{3n^{2}}\times \left( \frac{n\left( n\left( N+1\right)
+2N\right) }{4\pi }\right) ^{\frac{4n+2}{n+2}}.
\end{equation}%
Eqs.(26, 34) imply that%
\begin{equation}
\lambda ^{\frac{n-4}{n+2}}\leqslant \frac{1.45\times 10^{-10}n^{2}m_{pl}^{%
\frac{8-2n}{n+2}}}{M^{\frac{6}{n+2}}\left( \frac{n\left( n\left( N+1\right)
+2N\right) }{4\pi }\right) ^{\frac{4n+2}{n+2}}}
\end{equation}%
This is the condition for the brane tension according only to power spectrum
of the curvature perturbations observational value for a relevant $M$ energy
scale of the potential. For a complete study, it's necessary to check the
other inflationary parameters. According to Eq.(33), the inflationary
observables become 
\begin{eqnarray}
n_{s}-1 &=&-\frac{4n+2}{n\left( N+1\right) +2N}, \\
r &=&\frac{24n}{n\left( N+1\right) +2N}, \\
\frac{dn_{s}}{d\ln k} &=&-\frac{2\left( n+2\right) \left( 2n+1\right) }{%
\left( n\left( N+1\right) +2N\right) ^{2}}.
\end{eqnarray}%
Note that when $n\longrightarrow \infty ,$\ the inflation parameters $n_{s}$%
, $r$ and $\frac{dn_{s}}{d\ln k}$ reduce to expressions of exponential%
\textbf{\ }potential case(Eqs. 20, 21, 22). In the following, we plot these
observables as functions of $n$.\FRAME{dtbpFU}{4.2186in}{3.1073in}{0pt}{\Qcb{%
Fig.1: $n_{s}$ $vs$ $n$ for $N=50,55,60$ for monomial potential $V=M\protect%
\phi ^{n}$}}{}{fig1.eps}{\special{language "Scientific Word";type
"GRAPHIC";maintain-aspect-ratio TRUE;display "USEDEF";valid_file "F";width
4.2186in;height 3.1073in;depth 0pt;original-width 41.9045in;original-height
30.7917in;cropleft "0";croptop "1";cropright "1";cropbottom "0";filename
'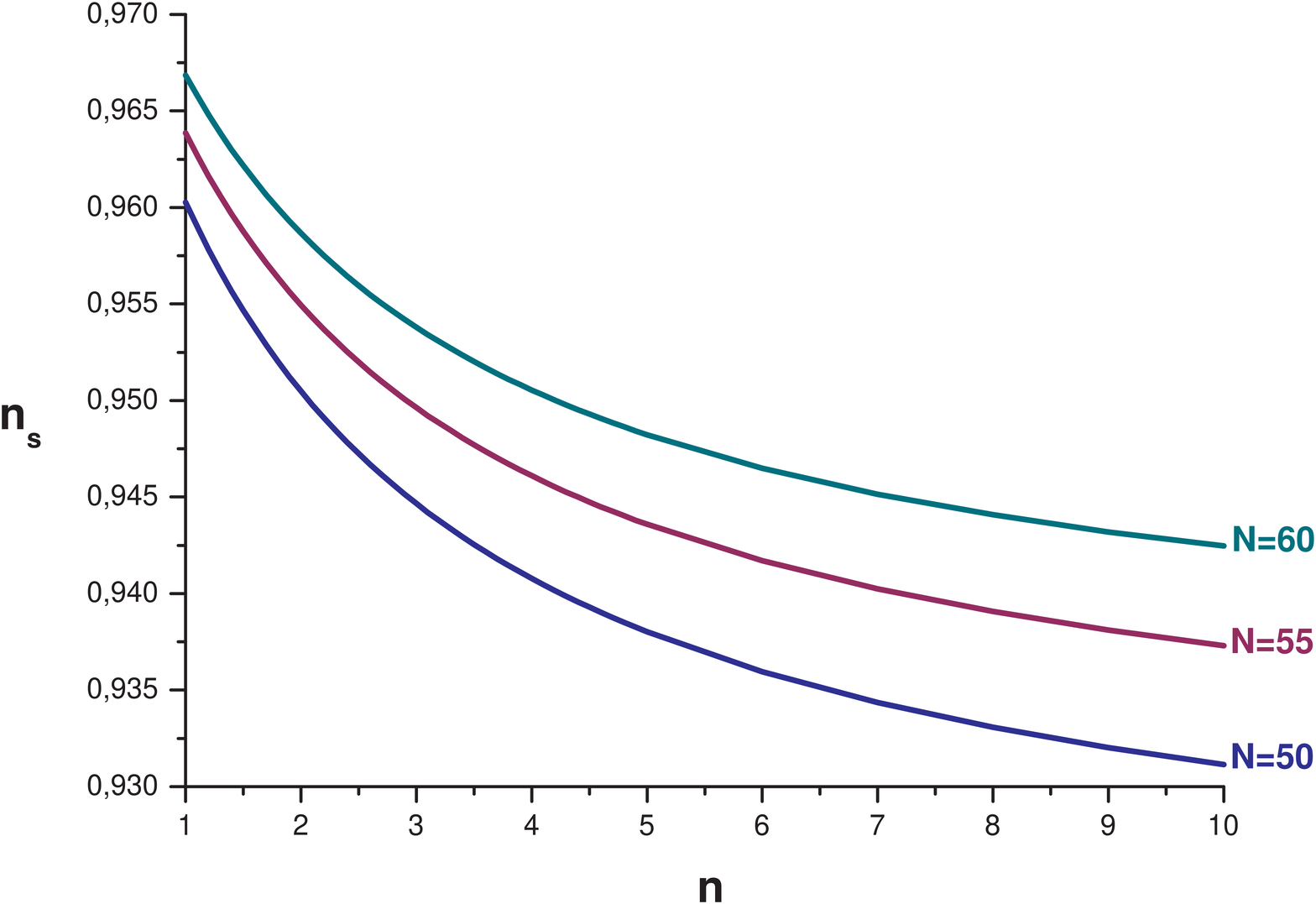';file-properties "XNPEU";}}The fig.1 shows that there exist a
significant region of variation of $n_{s}$ according to $n$, where the
results are consistent with observations, in particuler for large values of $%
N$ and small values of $n$. For large values of $n$, the scalar spectral
index becomes in disagreement with $WMAP7$ data except for very large $N$.
In that case, we recover the results obtained in the case of the exponential
potential (Eq.14).

In particular, for $n=2$ which corresponds to chaotic case, the
observational constraints for $n_{s}$ require that $N\geqslant 68$, Eq.(35)
reduces then to%
\begin{equation}
\lambda \gtrsim \frac{M^{3}}{2\times 10^{-27}m_{pl}^{2}}
\end{equation}%
\FRAME{dtbpFU}{4.0845in}{3.1073in}{0pt}{\Qcb{Fig.2: $r$ $vs$ $n$ for $%
N=50,55,60$ for monomial potential $V=M\protect\phi ^{n}$}}{}{fig2.eps}{%
\special{language "Scientific Word";type "GRAPHIC";maintain-aspect-ratio
TRUE;display "USEDEF";valid_file "F";width 4.0845in;height 3.1073in;depth
0pt;original-width 40.5735in;original-height 30.7968in;cropleft "0";croptop
"1";cropright "1";cropbottom "0";filename '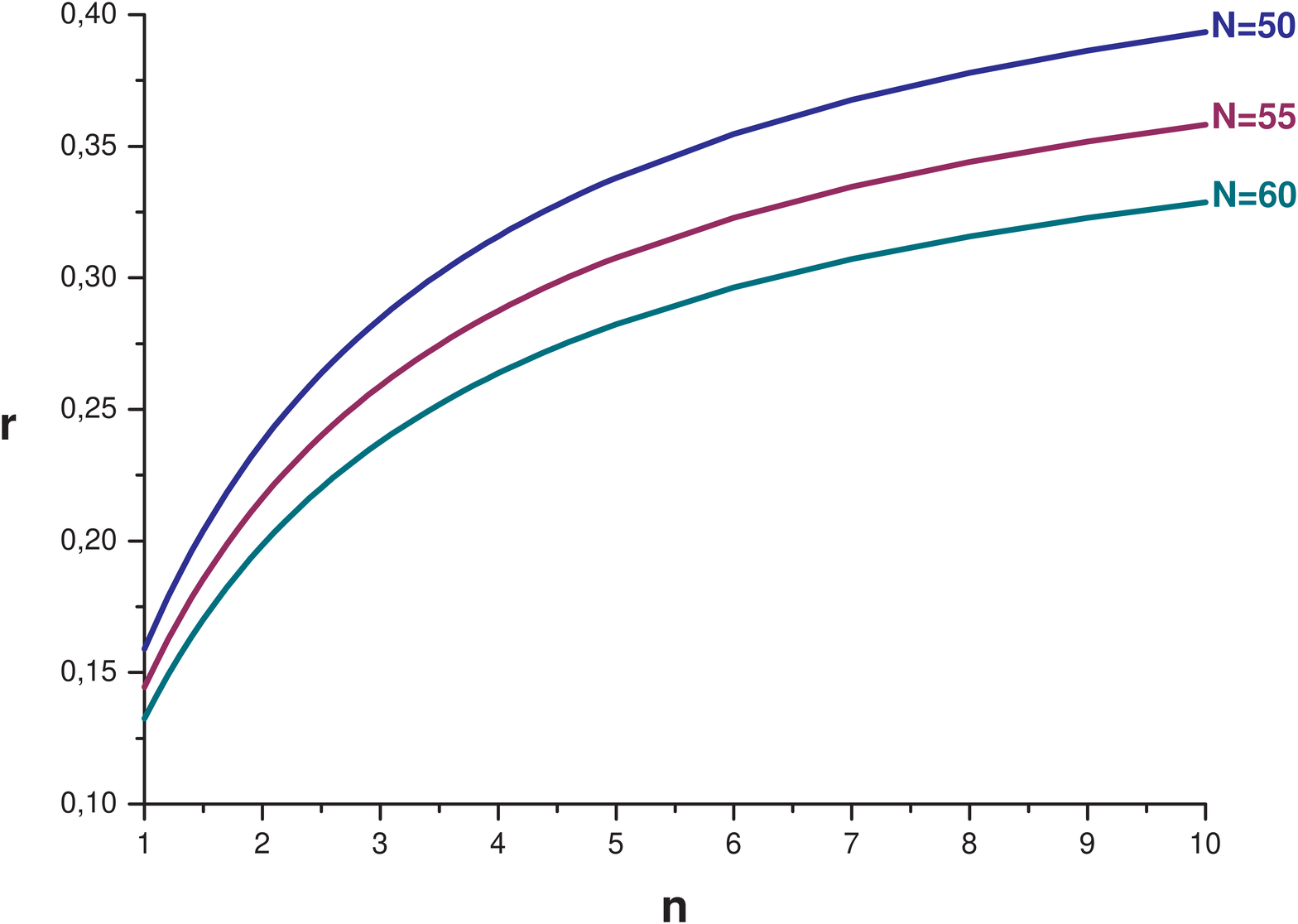';file-properties
"XNPEU";}}In this case, the fig.2 shows that we can recover the
observational results for large range of variation of exponent $n$ provided
that $N$ be large.\FRAME{dtbpFU}{4.3241in}{3.0588in}{0pt}{\Qcb{Fig.3: $\frac{%
dn_{s}}{dlnk}$ $vs$ $n$ for $N=50,55,60$ for monomial potential $V=M\protect%
\phi ^{n}$}}{}{fig3.eps}{\special{language "Scientific Word";type
"GRAPHIC";maintain-aspect-ratio TRUE;display "USEDEF";valid_file "F";width
4.3241in;height 3.0588in;depth 0pt;original-width 42.95in;original-height
30.303in;cropleft "0";croptop "1";cropright "1";cropbottom "0";filename
'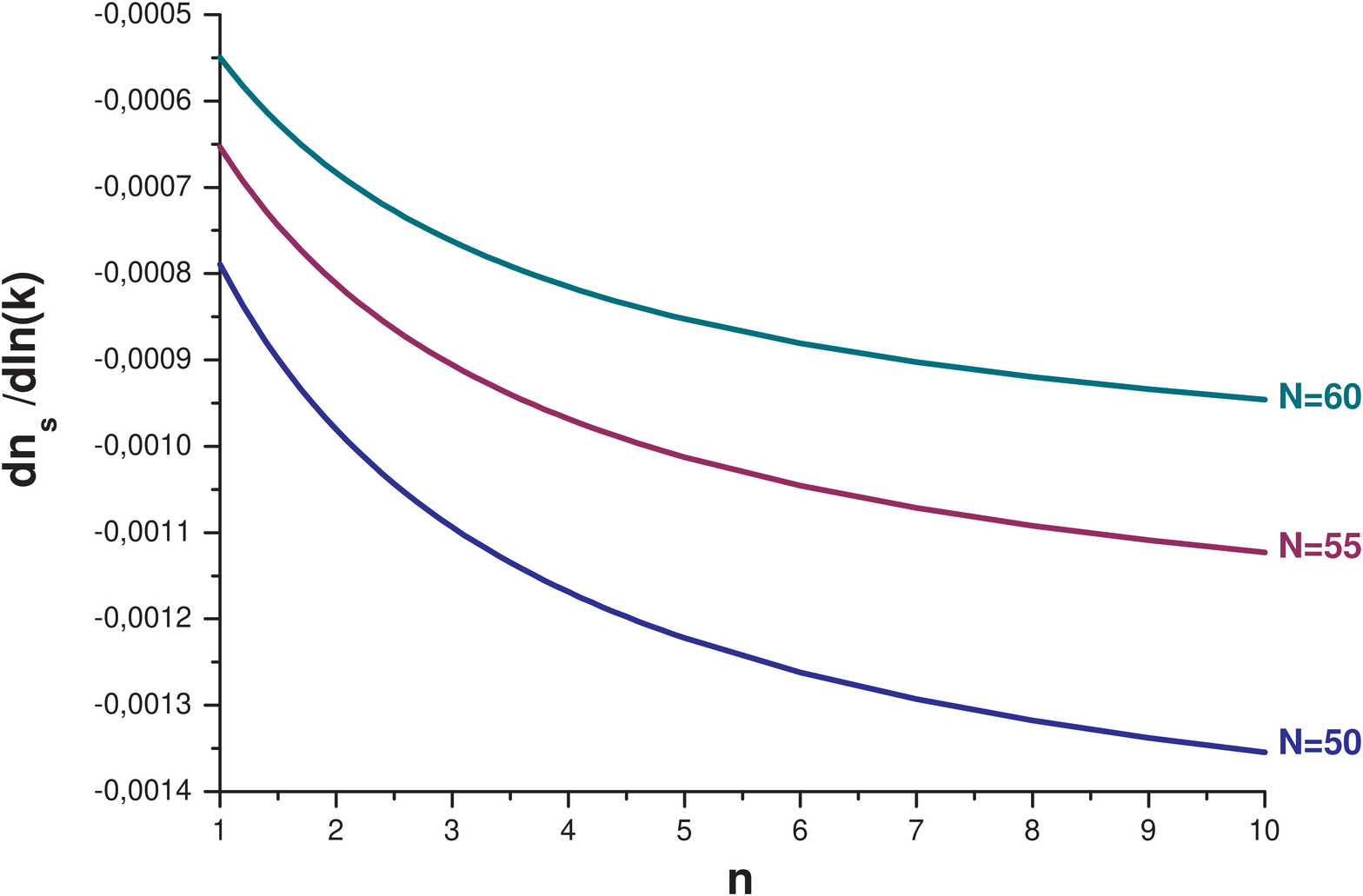';file-properties "XNPEU";}}In fig.3, we have ploted the variation
of $\frac{dn_{s}}{d\ln k}$ according to $n$. We see that the values of the
e-folding number $N$ compatibles with $WMAP7$ data, the running of the
scalar index is small and negative. Thus, we can conclude that the spectral
scalar index $n_{s}$ is scale invariant. This phenomena was already predicts
by inflation theory (Harrison-Zeldovich invariant)\cite{HZSI}.

\subsection{Inverse power law potential}

The inverse power law potential ($V=\frac{\mu }{\phi ^{m}}),$ has been
studied in various context, specially, for modelisation of quintessence
matter on the Braneworld inflation\cite{Lidsey} where it was shown that
inflation occurs only for $m>2$. This potential was also studied in
tachyonic inflation model\cite{L. Raul}. It have been shown that if the
tachyon dominates the background dynamics, then it will either go into a
dust-dominated phase ($m>2$), power-law expansion for $m=2$, or quasi
de-Sitter accelerated expansion for $m<<2$. We consider the following
inverse power law potential\cite{Bennai}%
\begin{equation}
V=\frac{\mu }{\phi ^{m}},
\end{equation}%
$\mu $ is a parameter of dimension $\left[ E\right] ^{4+m}.$

For this type of potential, the scalar spectral index and the ratio of
tensor to scalar perturbations take respectively the following expressions%
\begin{eqnarray}
n_{s}-1 &=&\left( -2m+1\right) \frac{\lambda m_{pl}^{2}}{2\pi }\frac{m}{\mu }%
\phi ^{\left( m-2\right) }, \\
r &=&\frac{6\lambda m_{pl}^{2}}{\pi }\frac{m^{2}}{\mu }\phi ^{\left(
m-2\right) }.
\end{eqnarray}%
The running of the scalar spectral index is presented by%
\begin{equation}
\frac{dn_{s}}{d\ln k}=-\frac{m_{pl}^{4}\lambda ^{2}m^{2}\left( m-2\right)
\left( 2m-1\right) }{8\pi ^{2}\mu ^{2}}\phi ^{2m-4}.
\end{equation}%
As above, we evaluate either the values of $\phi _{end}$ and $\phi _{\ast }$
using Eq.(13) 
\begin{eqnarray}
\phi _{end}^{\left( m-2\right) } &=&\frac{4\pi \mu }{\lambda m_{pl}^{2}m^{2}}%
, \\
\phi _{\ast }^{\left( m-2\right) } &=&\frac{4\pi \mu }{\lambda
m_{pl}^{2}m\left( m\left( N+1\right) -2N\right) }.
\end{eqnarray}%
The power spectrum of the curvature perturbations is given by%
\begin{equation}
P_{R}=\frac{16\pi \lambda ^{\frac{m+4}{m-2}}m_{pl}^{\frac{2m+8}{m-2}}}{%
3m^{2}\mu ^{\frac{6}{m-2}}}\times \left( \frac{4\pi }{m\left( m\left(
N+1\right) -2N\right) }\right) ^{\frac{2-4m}{m-2}}.
\end{equation}%
Eqs.(26, 46) imply that%
\begin{equation}
\lambda \geqslant \frac{\left( 1.27\times 10^{-10}m^{2}\right) ^{\frac{m-2}{%
m+4}}\mu ^{\frac{6}{m+4}}}{m_{pl}^{\frac{2m+8}{m+4}}\left( \frac{4\pi }{%
m\left( m\left( N+1\right) -2N\right) }\right) ^{\frac{2-4m}{m+4}}}.
\end{equation}%
As for the previous models, Eq.(47) shows the condition for the brane
tension according to power spectrum of the curvature perturbations
observational value for a relevant $\mu $ energy scale of the potential. To
complete our study, we analyse the variations of all other inflationary
parameters with respect to $m$.

So, in terms of $N$ and $m$, the inflationary parameters become%
\begin{eqnarray}
n_{s}-1 &=&\frac{2-4m}{m\left( N+1\right) -2N}, \\
r &=&\frac{24m}{m\left( N+1\right) -2N}, \\
\frac{dn_{s}}{d\ln k} &=&-\frac{2\left( m-2\right) \left( 2m-1\right) }{%
\left( m\left( N+1\right) -2N\right) ^{2}}.
\end{eqnarray}%
Note that when $m\longrightarrow \infty ,$\ the inflation parameters $n_{s}$%
, $r$ and $\frac{dn_{s}}{d\ln k}$ reduce to expressions of exponential%
\textbf{\ }potential case(Eqs. 20, 21, 22). In the following, we plot these
observables as functions of $n$.

\FRAME{dtbpFU}{4.1805in}{3.1073in}{0pt}{\Qcb{Fig.4: $n_{s}vs$ $m$ for $%
N=50,55,60$ for inverse powerl aw potential $V=\frac{\protect\mu }{\protect%
\phi ^{m}}$}}{}{fig4.eps}{\special{language "Scientific Word";type
"GRAPHIC";maintain-aspect-ratio TRUE;display "USEDEF";valid_file "F";width
4.1805in;height 3.1073in;depth 0pt;original-width 41.5283in;original-height
30.7917in;cropleft "0";croptop "1";cropright "1";cropbottom "0";filename
'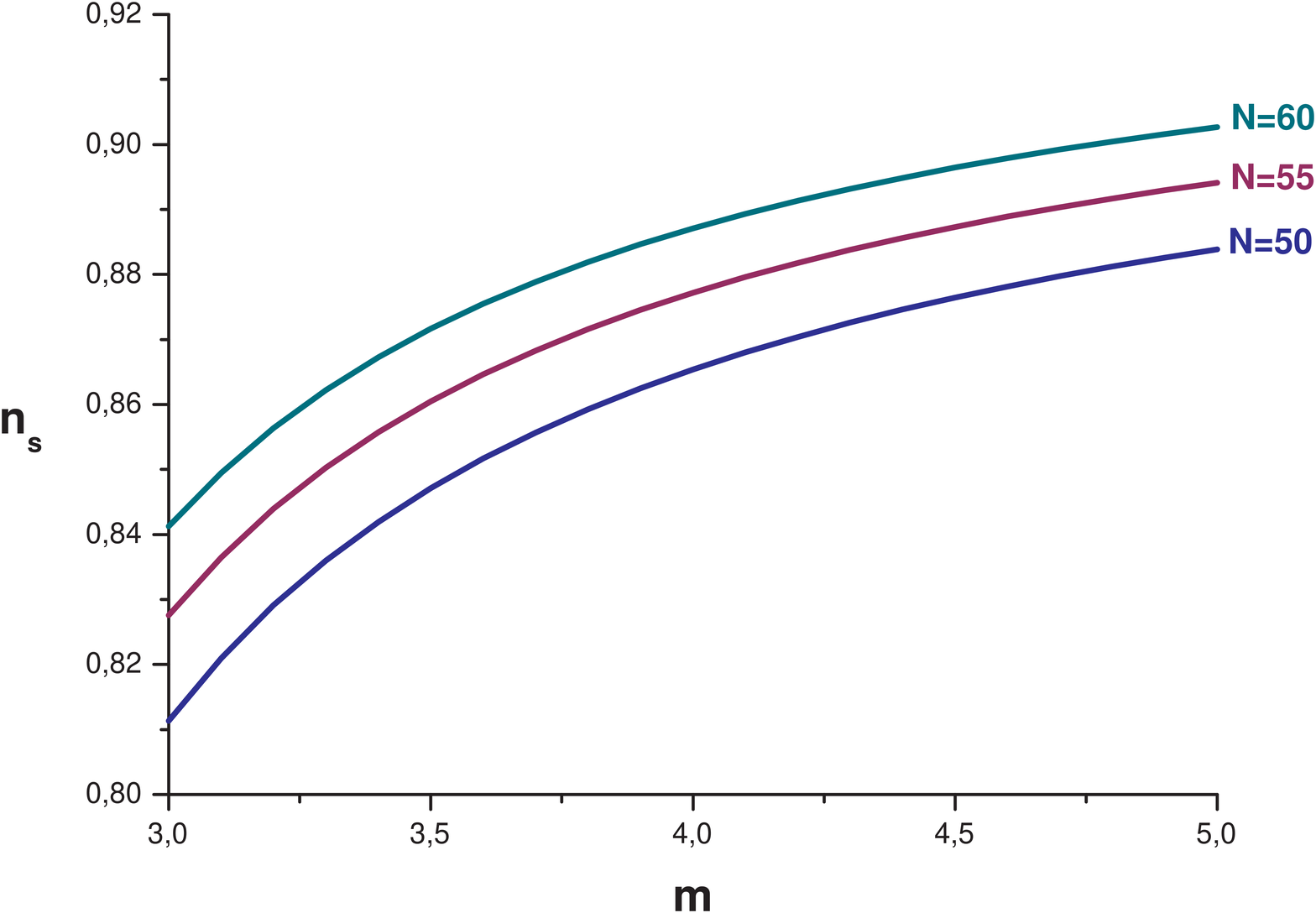';file-properties "XNPEU";}}Fig.4 shows the variation of $n_{s}$ as
function of $m.$ We observe that, in order to confront $n_{s}$ with WMAP7
data we must have very large value of $N$. This result is compatible with
the exponential potential case.\FRAME{dtbpFU}{4.0413in}{3.1081in}{0pt}{\Qcb{%
Fig.5: $r$ $vs$ $m$ for $N=50,55,60$ for inverse power law potential $V=%
\frac{\protect\mu }{\protect\phi ^{m}}$}}{}{fig5.eps}{\special{language
"Scientific Word";type "GRAPHIC";maintain-aspect-ratio TRUE;display
"USEDEF";valid_file "F";width 4.0413in;height 3.1081in;depth
0pt;original-width 40.1429in;original-height 30.8098in;cropleft "0";croptop
"1";cropright "1";cropbottom "0";filename '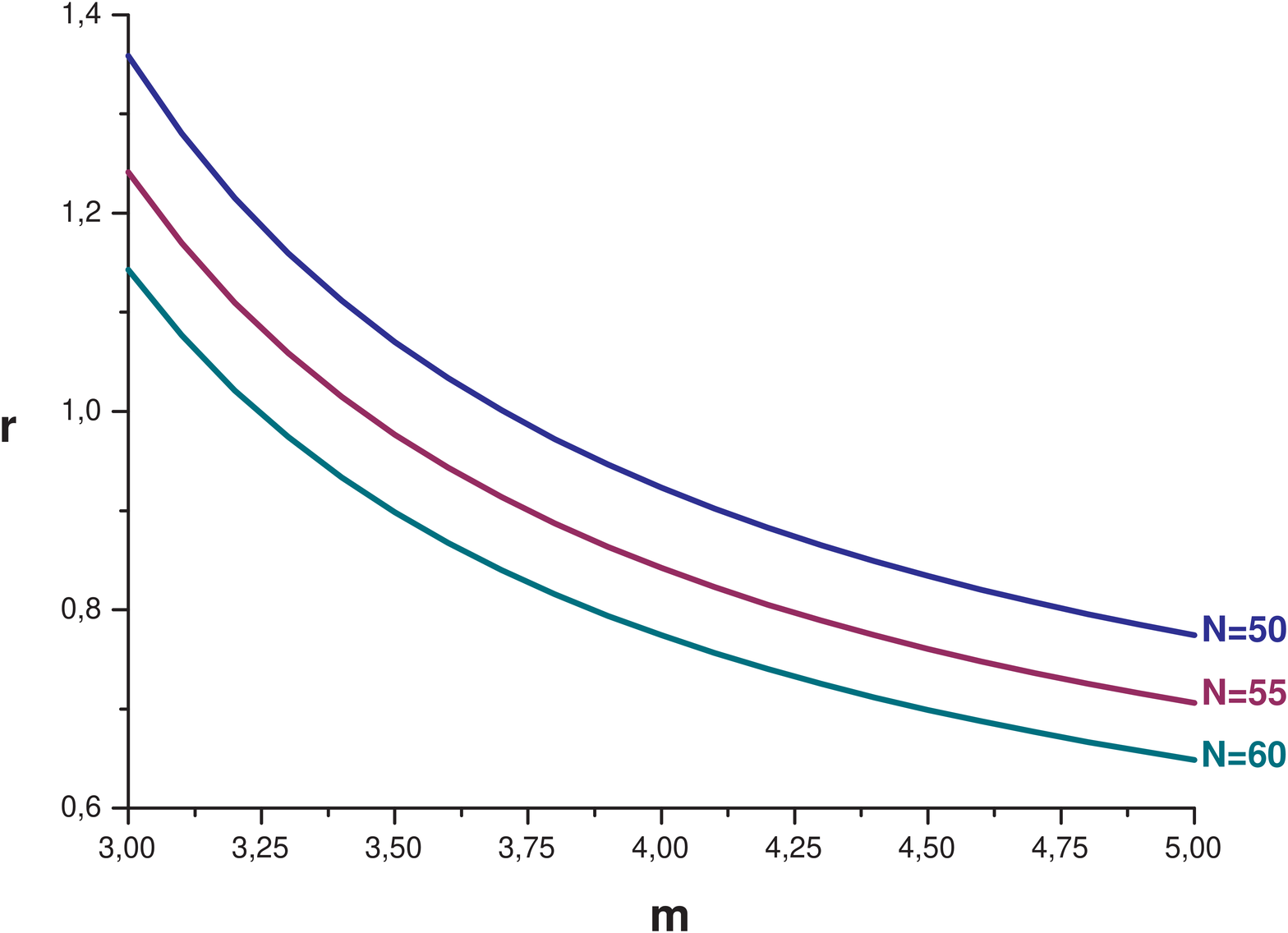';file-properties
"XNPEU";}}This figure which presents $r$ as function of $m$, confirme our
above results concerning the very large values of e-folding number $N$ as
condition of inflation.

In particular, for $m=10$ , the observational constraints for $n_{s}$ and $r$
require that $N\geqslant 128.$ Then, for $m=10$ and $N=128$, Eq.(47) reduces
to%
\begin{equation}
\lambda \gtrsim \frac{3.75\times 10^{-13}\mu ^{\frac{3}{7}}}{m_{pl}^{2}}
\end{equation}%
\FRAME{dtbpFU}{4.3249in}{3.0891in}{0pt}{\Qcb{Fig.6: $\frac{dn_{s}}{dlnk}$ $vs
$ $m$ for $N=50,55,60$ for inverse power law potential $V=\frac{\protect\mu 
}{\protect\phi ^{m}}$}}{}{fig6.eps}{\special{language "Scientific Word";type
"GRAPHIC";maintain-aspect-ratio TRUE;display "USEDEF";valid_file "F";width
4.3249in;height 3.0891in;depth 0pt;original-width 42.963in;original-height
30.6109in;cropleft "0";croptop "1";cropright "1";cropbottom "0";filename
'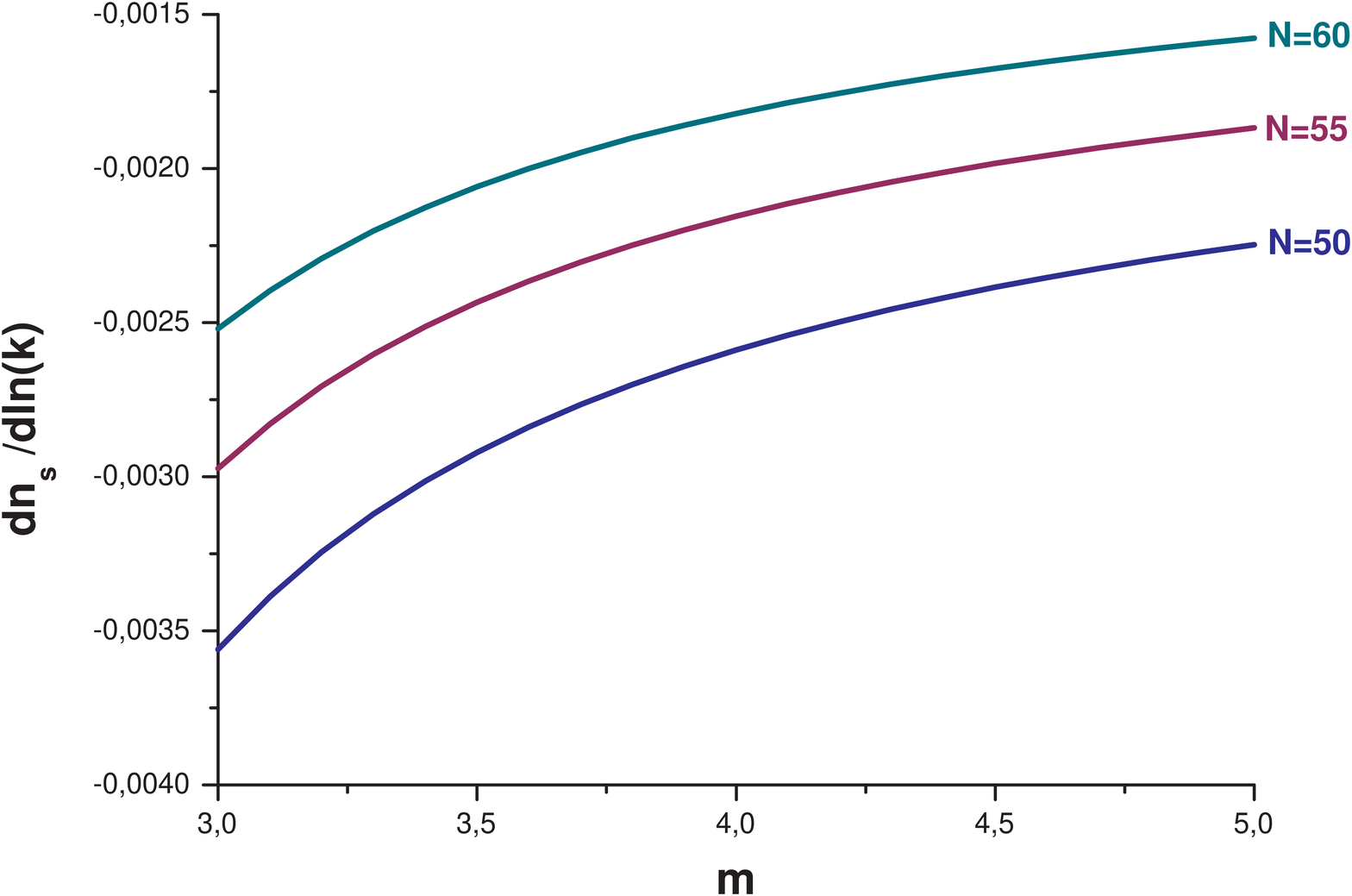';file-properties "XNPEU";}}It's clear, as shown in fig.6, that the
running of the scalar index $\frac{dn_{s}}{d\ln k}$ becomes negligible for
very large $N$, as predicted by WMAP7 observations.

\subsection{Recapitulate}

We present in this subsection a summary concerning the different
inflationary parameters for the three types of potential at $V=V_{\ast }$
and $\phi =\phi _{\ast }$.

\begin{center}
\begin{tabular}[t]{|l|c|c|c|}
\hline
\emph{Potentials} & Monomial $V=M\phi ^{n}$ & Inverse power law $V=\frac{\mu 
}{\phi ^{m}}$ & Exponential $V=V_{0}\exp \left( -\frac{\alpha }{m_{p}}\phi
\right) $ \\ \hline
\multicolumn{1}{|c|}{$\epsilon $} & $\frac{n}{n\left( N+1\right) +2N}$ & $%
\frac{m}{m\left( N+1\right) -2N}$ & $\frac{1}{N+1}$ \\ \hline
\multicolumn{1}{|c|}{$\eta $} & $\frac{n-1}{n\left( N+1\right) +2N}$ & $%
\frac{m+1}{m\left( N+1\right) -2N}$ & $\frac{1}{N+1}$ \\ \hline
\multicolumn{1}{|c|}{$n_{s}-1$} & $-\frac{4n+2}{n\left( N+1\right) +2N}$ & $%
\frac{2-4m}{m\left( N+1\right) -2N}$ & $-\frac{4}{N+1}$ \\ \hline
\multicolumn{1}{|c|}{$r$} & $\frac{24n}{n\left( N+1\right) +2N}$ & $\frac{24m%
}{m\left( N+1\right) -2N}$ & $\frac{24}{N+1}$ \\ \hline
\multicolumn{1}{|c|}{$\frac{dn_{s}}{d\ln k}$} & $-\frac{2\left( n+2\right)
\left( 2n+1\right) }{\left( n\left( N+1\right) +2N\right) ^{2}}$ & $-\frac{%
2\left( m-2\right) \left( 2m-1\right) }{\left( m\left( N+1\right) -2N\right)
^{2}}$ & $-\frac{4}{\left( N+1\right) ^{2}}$ \\ \hline
\end{tabular}
\end{center}

This table shows that, for $V=V_{\ast }$ and $\phi =\phi _{\ast },$ all
inflationary parameters depend only on the e-folding number $N$ and
exponents of the scalar fields for the two first potentials. Note that for
large $n$ and $m$, the monomial and inverse power law potentials lead to
similar results in the exponential potential case.

\section{Conclusion}

In this work, we have studied various types of potentials by considering
different inflationary perturbation spectrum paramaters. We have analyzed in
particular the exponential potential: $V=V_{0}\exp \left( -\frac{\alpha }{%
m_{p}}\phi \right) $, monomial potential: $V=M\phi ^{n}$ and inverse power
law model: $V=\frac{\mu }{\phi ^{m}}$ in the framework of the \emph{%
Randall-Sundrum} type-2 model. We have shown that, the monomial potential
provides the best fit results with recent $WMAP7$ observational constraints.
Indeed, for the exponential potential, the observational constraints require
that $N>107,$ so that $n_{s}$ and $r$ simultaneously are consistent with $%
WMAP7$ data assuming a negligible running. However, for the inverse power
law potential, the observational constraints on the inflationary parameters $%
n_{s}$ and $r$ have lead to consider a very large e-folding number $N$.

On the other hand, for the monomial potential, the parameters $n_{s}$ and $r$
are consistents with the observations for small values of $n$ and large $N$
(notably for the scalar spectral index $n_{s}$). Finally, note that, as $n$
and $m$ increase, the inverse power law potential and monomial present the
some behaviors as the exponential one.

\end{document}